\documentclass[twocolumn,twoside,slac_two]{revtex4}
\usepackage{graphicx}
\usepackage{fancyhdr}
\begin{document}

\title{Neutrino Theory: Some Recent Developments}

%

\author{Ernest Ma}
\affiliation{Department of Physics and Astronomy, University of California, 
Riverside, CA 92521, USA}

\begin{abstract}
Selected topics in neutrino theory are discussed, including tribimaximal 
mixing and its $A_4$ realization, variations of the seesaw mechanism 
allowing for observable deviations from unitarity of the neutrino mixing 
matrix, and the radiative generation of neutrino mass from dark matter.
\end{abstract}

\maketitle

\thispagestyle{fancy}


\section{Introduction}
In this talk, I will discuss a number of developments in neutrino theory over 
the past three or four years.  I will not have time to cover many interesting 
topics which are being actively pursued, but I will mention some in passing. 
In the following, I will give a brief account of the history of neutrino 
tribimaximal mixing and the tetrahedral symmetry $A_4$ which explains it. 
I will then discuss variations of the seesaw mechanism, with focus on 
the possibility that it may be testable at the TeV scale through the 
nonunitarity of the neutrino mxing matrix.  The possible connection 
between the radiative seesaw mechanism and dark matter, i.e.{\it scotogenic} 
neutrino mass, is then explored with three examples.

\section{Brief History of Neutrino Tribimaximal Mixing}
In 1978, soon after the appearance of hints for the existence of the third 
family of quarks and leptons, it was conjectured independently by 
Cabibbo~\cite{c78} and Wolfenstein~\cite{w78} that the mixing matrix linking 
charged leptons to neutrinos could be given by
\begin{equation}
U^{CW}_{l \nu} = {1 \over \sqrt{3}} \pmatrix{1 & 1 & 1 \cr 1 & \omega & 
\omega^2 \cr 1 & \omega^2 & \omega},
\end{equation}
where $\omega = \exp(2 \pi i/3) = -1/2 + i \sqrt{3}/2$.  In the conventional 
notation, this corresponds to
\begin{equation}
\theta_{23} = \theta_{12} = 45^\circ, ~~~ \theta_{13} = 35.3^\circ, ~~~
\delta_{CP} = 90^\circ.
\end{equation}
It should dispel the myth that everybody expected small mixing angles in the 
lepton sector as in the quark sector.

In 2002, after neutrino oscillations have been established, Harrison, Perkins, 
and Scott~\cite{hps02} proposed the tribimaximal mixing matrix, i.e.
\begin{equation}
U^{HPS}_{l \nu} = \pmatrix{\sqrt{2/3} & 1/\sqrt{3} & 0 \cr -1/\sqrt{6} & 
1/\sqrt{3} & -1/\sqrt{2} \cr -1/\sqrt{6} & 1/\sqrt{3} & 1/\sqrt{2}} \sim 
(\eta_8, ~\eta_1, ~\pi^0),
\end{equation}
where the columns are evocative of members of the meson nonet in terms of 
quark-antiquark bound states.

In 2004, I discovered~\cite{m04} the simple connection:
\begin{equation}
U^{HPS}_{l \nu} = (U^{CW}_{l \nu})^\dagger \pmatrix{1 & 0 & 0 \cr 0 & 1/\sqrt{2} & 
-1/\sqrt{2} \cr 0 & 1/\sqrt{2} & 1/\sqrt{2}} \pmatrix{0 & 1 & 0 \cr 1 & 0 & 0 
\cr 0 & 0 & i},
\end{equation}
where the last matrix merely redefines $\nu_{1,2,3}$ to agree with the usual 
convention.  This shows that $U^{HPS}_{l \nu}$ is obtained from the mismatch 
of $U^{CW}_{l \nu}$ and a simple $45^\circ$ rotation in the $2-3$ sector. 
It means that if the charged-lepton mass matrix is given by
\begin{equation} 
{\cal M}_l = U^{CW}_{l \nu} \pmatrix{m_e & 0 & 0 \cr 0 & m_\mu & 0 \cr 0 & 0 & 
m_\tau} (U^l_R)^\dagger,
\end{equation}
and the Majorana neutrino mass matrix has $2-3$ reflection symmetry, with 
zero $1-2$ and $1-3$ mixing, i.e.
\begin{equation}
{\cal M}_\nu = \pmatrix{a+2b & 0 & 0 \cr 0 & a-b & d \cr 0 & d & a-b},
\end{equation}
$U^{HPS}_{l \nu}$ will be obtained, but {\it how}? Tribimaximal mixing means:
\begin{equation}
\theta_{13}=0, ~~~ \sin^2 2 \theta_{23}=1, ~~~ \tan^2 \theta_{12} = 1/2.
\end{equation}
In 2002, when HPS proposed it, world data were not precise enough to test 
this hypothesis.  In 2004, when I derived it, SNO data implied 
$\tan^2 \theta_{12} = 0.40 \pm 0.05$, which was not so encouraging. 
Then in 2005, revised SNO data obtained $\tan^2 \theta_{12} = 0.45 \pm 0.05$, 
and tribimaximal became a household word, unleashing a glut of papers. 
At present, the central value of $\tan^2 \theta_{12}$ is more like $0.47$, 
making it even closer to 1/2.

\section{Tetrahedral Symmetry A$_4$}
\subsection{What is A$_4$?}
For three families, we should look for a group with an irreducible 
\underline{3} representation, the simplest of which is A$_4$, the group of 
the even permutation of four objects. It has 12 elements and 4 equivalence 
classes with 3 inequivalent one-dimensional represenations and 1 
three-dimensional one. As such, it is tailor-made for describing 3 familes. 
Its character table is given below.

\begin{table}[htb]
\begin{center}
\caption{Character Table of A$_4$.}
\begin{tabular}{|c|c|c|c|c|c|c|}
\hline class & $n$ & $h$ & $\chi_1$ & $\chi_{1'}$ & $\chi_{1''}$ & $\chi_3$ \\ 
\hline
 $C_1$ & 1 & 1 & 1 & 1 & 1 & 3 \\
 $C_2$ & 4 & 3 & 1 & $\omega$ & $\omega^2$ & 0 \\
 $C_3$ & 4 & 3 & 1 & $\omega^2$ & $\omega$ & 0 \\
 $C_4$ & 3 & 2 & 1 & 1 & 1 & $-1$ \\
\hline
\end{tabular}
\label{a4}
\end{center}
\end{table}
The cube root of unity, i.e. $\omega$, appears again, and you can catch a 
glimpse of $U^{CW}_{l \nu}$ in the above.  The key to how this leads to 
Eq.~(5) is the multiplication rule:
\begin{eqnarray}
\underline{3} \times \underline{3} &=& \underline{1} (11+22+33) \nonumber \\ 
&+& \underline{1}' (11 + \omega^2 22 + \omega 33) + 
\underline{1}'' (11 + \omega 22 + \omega^2 33) \nonumber \\ 
&+& \underline{3} (23, 31, 12) + \underline{3} (32, 13, 21).
\end{eqnarray}
Note that $\underline{3} \times \underline{3} \times \underline{3} = 
\underline{1}$ is possible in A$_4$, i.e. $a_1 b_2 c_3 +$ permutations.   

The non-Abelian finite group A$_4$ is also the symmetry group of the regular 
tetrahedron, one of five perfect three-dimensional geometric solids, as shown 
below~\cite{m02}.

\begin{table}[htb]
\begin{center}
\caption{Perfect Three-Dimensional Geometric Solids.}
\begin{tabular}{|c|c|c|c|c|}
\hline solid & faces & vertices & Plato & group \\ 
\hline
tetrahedron & 4 & 4 & fire & A$_4$ \\
octahedron & 8 & 6 & air & S$_4$ \\
cube & 6 & 8 & earth & S$_4$ \\
icosahedron & 20 & 12 & water & A$_5$ \\
dodecahedron & 12 & 20 & quintessence & A$_5$ \\
\hline
\end{tabular}
\label{perfect}
\end{center}
\end{table}
 
Most models of neutrino tribimaximal mixing are based on A$_4$.  Some use 
S$_4$ and recently, A$_5$ has also been proposed~\cite{es09}.  It is amusing 
to note the parallel case of five string theories in ten dimensions: Type I 
is dual to Heterotic SO(32), Type IIA is dual to Heterotic $E_8 \times E_8$, 
and Type IIB is self-dual.

\subsection{Two Steps to Tribimaximal Mixing}
To realize neutrino tribimaximal mixing using A$_4$, there are two steps. 
The first is to obtain Eq.~(5). Here there are two options. (A) Let 
$(\nu_i,l_i) \sim \underline{3}$, $l^c_i \sim \underline{1}, \underline{1}', 
\underline{1}''$, and $(\phi^0_i,\phi^-_i) \sim \underline{3}$ with 
$v_1=v_2=v_3$.  This was the original proposal of Ma and 
Rajasekaran~\cite{mr01}.  (B) Let $(\nu_i,l_i), l^c_i \sim \underline{3}$, 
and $(\phi^0_i,\phi^-_i) \sim \underline{1}, \underline{3}$, with 
$v_1=v_2=v_3$.  This choice~\cite{m06-1} enables A$_4$ to be compatible with 
grand unified theories where quarks and leptons belong to a single multiplet, 
such as the \underline{16} of SO(10).  In either case, the diagonalization 
of ${\cal M}_l$ yields $U^{CW}_{l \nu}$ automatically, with arbitrary 
$m_e,m_\mu,m_\tau$. This is the crucial reason for the success of A$_4$.
Note that A$_4$ breaks to the residual symmetry Z$_3$ which maintains the 
condition $v_1=v_2=v_3$.

The second step is to obtain Eq.~(6).  The most straightforward way is to use 
Higgs triplets $(\xi^{++}_i,\xi^+_i,\xi^0_i) \sim \underline{1}, \underline{3}$ 
with $u_2=u_3=0$, resulting in~\cite{m04,af05}
\begin{equation}
{\cal M}_\nu = \pmatrix{a & 0 & 0 \cr 0 & a & d \cr 0 & d & a}.
\end{equation}
This is the simplest realization of tribimaximal mixing, with neutrino mass 
eigenvalues $a+d$, $a$, $-a+d$, allowing only normal hierarchy~\cite{m05}. 
Here A$_4$ breaks to the residual symmetry Z$_2$ which maintains the 
condition $u_2=u_3=0$.

To summarize, the keys to neutrino tribimaximal mixing are (1) the choice of 
symmetry, i.e. A$_4$ or S$_4$, etc., (2) the choice of lepton and Higgs 
representations, (3) residual symmetries Z$_3$ and Z$_2$ in different sectors. 
The last condition is the most technically challenging.  In a complete model,  
a large number of extra particles and auxiliary symmetries are often required, 
sometimes also with nonrenormalizable operators and perhaps even in the 
context of extra space dimensions.
{\it The above is an example of how group theory alone could determine mixing 
angles, but leaves all masses free.}  Applying this idea to the quark sector, 
it has been shown~\cite{l07,bhl08} that the Cabibbo angle = $\pi/14$ could 
come from D$_7$ breaking to Z$_2$.

\section{Seesaw Variations and the Nonunitarity of the Neutrino Mixing Matrix}
\subsection{Canonical Seesaw}
With one doublet neutrino $\nu$ and one singlet neutrino $N$, their 
$2 \times 2$ mass matrix is the well-known
\begin{equation}
{\cal M}_{\nu N} = \pmatrix{0 & m_D \cr m_D & m_N},
\end{equation}
resulting in the famous seesaw formula $m_\nu \simeq -m_D^2/m_N$. Hence 
$\nu-N$ mixing $\simeq m_D/m_D \simeq \sqrt{m_\nu/m_N} < 10^{-6}$, for 
$m_\nu < 1$ eV and $m_N > 1$ TeV.  This means that even if $N$ is light 
enough to be kinematically accessible, it cannot be produced at the Large 
Hadron Collider (LHC) because its interaction with other particles is 
too weak.  I assume here that $m_N$ is not much below the electroweak 
symmetry breaking scale of $10^2$ GeV.  If this assumption is relaxed, 
there could be interesting phenomenological implications~\cite{ahpz09,ghj09}.

\subsection{Inverse Seesaw}
Consider next one $\nu$ and two singlets $N_{1,2}$. Their $3 \times 3$ mass 
matrix is then
\begin{equation}
{\cal M}_{\nu N} = \pmatrix{0 & m_D & 0 \cr m_D & m_1 & m_N \cr 0 & m_N & m_2},
\end{equation}
resulting in $m_\nu \simeq m_D^2 m_2/m_N^2$.  Since the limit $m_1=m_2=0$ 
corresponds to lepton number conservation ($L=1$ for $\nu$ and $N_2$, $L=-1$ 
for $N_1$), their smallness is natural.  Thus $m_\nu$ is small, not because 
$m_N$ is big, but rather that $m_2$ is small.  This is the inverse seesaw 
mechanism~\cite{ww83,mv86,m87}.  Here $\nu-N_1$ mixing remains small 
as in the previous case, but $\nu-N_2$ mixing $\simeq m_D/m_N$ may now be big, 
say of order $10^{-2}$.  Thus $U_{l \nu}$ may deviate significantly from 
being unitary and the effect may be decipherable~\cite{m09-1} at the LHC.  
If all three neutrinos are considered, this will also be a new source of 
lepton flavor violation, which may also be experimentally observable.

\subsection{Seesaw Textures}
For two families, consider $\nu_{1,2}$ and $N_{1,2}$, with the special 
choice~\cite{bw90}
\begin{equation}
{\cal M}_{\nu N} = \pmatrix{0 & 0 & a_1 b_1 & a_1 b_2 \cr 0 & 0 & a_2 b_1 & a_2 
b_2 \cr a_1 b_1 & a_2 b_1 & M'_1 & 0 \cr a_1 b_2 & a_2 b_2 & 0 & M'_2}.
\end{equation}
In that case, the arbitrary imposed condition
\begin{equation}
{b_1^2 \over M'_1} + {b_2^2 \over M'_2} = 0
\end{equation}
renders all two light neutrinos massless, and yet $\nu-N$ mixing may be 
large, because $a_i b_j$ need not be small.  Small deviations from this 
texture will allow small neutrino masses to appear, but retain the large 
$\nu-N$ mixing.  This is an active topic of study with many recent papers, 
including Refs.~\cite{p05,ks07}.

To understand the mechanism and symmetry of the texture hypothesis, change 
the neutrino basis to~\cite{hm09}
\begin{equation}
{\cal M}_{\nu N} = \pmatrix{0 & 0 & m_1 & 0 \cr 0 & 0 & 0 & m_2 \cr m_1 & 0 
& M_1 & M_3 \cr 0 & m_2 & M_3 & M_2}.
\end{equation}
Then Eq.~(13) implies $m_1=M_1=0$, so that $\nu_1$ and $\nu'_2 = (M_3 \nu_2 
- m_2 N_1)/\sqrt{M_3^2+m_2^2}$ are massless, showing how large mixing actually 
occurs, i.e. through the {\it inverse} seesaw mechanism.  However, lepton 
number conservation would not only forbid $M_1$ but also $M_2$, which is 
in fact aribitrary here.  Where is the symmetry which does this?

Let $\nu_{1,2},N_{1,2}$ have $L=1,1,3,-1$.  Add the usual Higgs doublet 
$(\phi_1^+,\phi_1^0)$ with $L=0$ and the Higgs singlet $\chi_2$ with $L=2$. 
Then $m_2$ comes from $\langle \phi_1^0 \rangle$, $M_2$ from $\langle \chi_2 
\rangle$, and $M_3$ from $\langle \chi_2^\dagger \rangle$.  The absence of 
a Higgs doublet with $L=-4$ and a singlet with $L=\pm6$ means that $m_1=M_1=0$ 
at tree level.  However, $M_1$ is induced in one loop, from the coupling of 
$N_1$ to $N_2$ through $\chi_2$.  This effect is proportional to $M_2$ and 
finite, because the breaking of $L$ by $\langle \chi_2 \rangle$ results in 
Re$(\chi_2)$ and Im$(\chi_2)$ having different masses, the latter being zero 
if $L$ is broken spontaneously and nonzero if it is also broken explicitly, 
by a soft term such as $\chi_2^2$ for example.  Note that $(-)^L$ is still 
conserved.  Thus $\nu'_2$ acquires an inverse seesaw mass $M_1 m_2^2/M_3^2$. 
Once $\nu'_2$ is massive, $\nu_1$ also gets a two-loop radiative 
mass~\cite{bm88} from the exchange of two $W$ bosons.

\subsection{Seesaw Extensions}
Type I seesaw, i.e. using the singlet femion $N$, is generically difficult 
to verify, even if $m_N$ is at the TeV scale.  If large $\nu-N$ mixing 
occurs, as discussed in the above, then there is a chance.  On the other hand, 
the Higgs triplet $(\xi^{++},\xi^+,\xi^0)$ of Type II seesaw and the fermion 
triplet $(\Sigma^+,\Sigma^0,\Sigma^-)$ of Type III seesaw have accompanying 
charged particles as well as electroweak gauge interactions which are much 
easier to find at the LHC.  These are also active topics of study with many 
recent papers, including Ref.~\cite{aa09}.  Another attractive possibility 
is to gauge lepton number or some related quantity, such as $B-L$.  This 
would predict a $Z'$ boson which couples to $N$ and allows the latter to be 
produced, resulting thus in many interesting phenomenological 
consequences~\cite{k08,hkor08,ko09,adrs09,kkr09,bbms09,bmt09}.
However, $U(1)_{B-L}$ is not orthogonal to $U(1)_Y$.  It is theoretically 
much better to use $U(1)_\chi$ instead~\cite{m09-2,fhl09}, because it comes 
from the breaking of $SO(10) \to SU(5) \times U(1)_\chi \to SU(3)_C \times 
SU(2)_L \times U(1)_Y \times U(1)_\chi$ through the vacuum expectation value of 
the Higgs \underline{45} along the $(\underline{24},0)$ component.  Here 
$Q_\chi = 5(B-L)- 4Y$.

\section{Radiative Seesaw and Dark Matter: Scotogenic Neuttrino Mass}
\subsection{One-Loop Prototype Model}
Neutrino mass may be radiative in origin.  This is a very old idea, but 
if the particles in the loop are all odd under an exactly conserved Z$_2$ 
symmetry whereas the ordinary particles are even, then a connection with 
dark matter may be established.  The simplest such model~\cite{m06-2} 
assigns $N_{1,2,3}$ and a second scalar doublet~\cite{dm78} $(\eta^+,\eta^0)$ 
to be odd under Z$_2$.  Hence $\nu N \phi^0$ is forbidden and $\nu N \eta^0$ 
is allowed, but $\langle \eta^0 \rangle = 0$, so that $N$ is not the 
Dirac mass partner of $\nu$.  A finite neutrino mass is generated in one 
loop because Re$(\eta^0)$ and Im$(\eta^0)$ have different masses from the 
allowed quartic scalar interaction term $(\lambda_5/2)(\Phi^\dagger \eta)^2 
+ H.c.$  This mass splitting also enabless Re$(\eta^0)$ or Im$(\eta^0)$ to 
be a realistic dark-matter candidate, as studied in some detail two months 
later by Barbieri, Hall, and Rychkov~\cite{bhr06}.  They call $\eta$ the 
inert Higgs doublet, but it is neither inert because it has gauge 
interactions, nor a ``Higgs'' because it has no vacuum expectation value. 
I call it the {\it dark scalar doublet}. 

The one-loop radiative seesaw neutrino mass is easily calculated:
\begin{equation}
({\cal M}_\nu)_{\alpha \beta} = \sum_i {h_{\alpha i} h_{\beta i} M_i \over 16 \pi^2} 
\left[ f \left( {m_R^2 \over M_i^2} \right) - f \left( {m_I^2 \over M_i^2} 
\right) \right],
\end{equation}
where $f(x) = -x \ln x/(1-x)$.  Let $m_R^2 - m_I^2 = 2 \lambda_5 v^2 << 
m_0^2 = (m_R^2 + m_I^2)/2$, then
\begin{equation}
({\cal M}_\nu)_{\alpha \beta} = \sum_i {h_{\alpha i} h_{\beta i} \over M_i} 
I \left( {m_0^2 \over M_i^2} \right),
\end{equation}
where
\begin{equation}
I(x) = {-\lambda_5 v^2 \over 8 \pi^2} \left( {1 \over 1-x} \right) \left[ 
1 + {\ln x \over 1-x} \right].
\end{equation}
For $x_i << 1$, i.e. $N_i$ very heavy, this reduces to
\begin{equation}
({\cal M}_\nu)_{\alpha \beta} = {-\lambda_5 v^2 \over 8 \pi^2} \sum_i 
{h_{\alpha i} h_{\beta i} \over M_i} (1 + \ln x_i)
\end{equation}
instead of the canonical seesaw expression of $v^2 \sum_i h_{\alpha i} h_{\beta i} 
/M_i$. In the context of leptogenesis, if this mechanism is 
extended~\cite{m06-3} to include supersymmetry, the Davidson-Ibarra lower 
bound~\cite{di02} of about $10^9$ GeV for the lightest $N_i$ may then be 
evaded, avoiding thus a potential conflict of gravitino overproduction 
and thermal leptogenesis. 

\subsection{Supersymmetric SU(5) Completion}
This model of scotogenic neutrino mass also has a straightforward SU(5) 
completion~\cite{m08} with gauge-coupling unification.  Simply add the 
complete superfield multiplets
\begin{eqnarray}
\underline{5} &=& h(3,1,-1/3) + (\eta_2^+,\eta_2^0)(1,2,1/2), \\ 
\underline{5}^* &=& h^c(3^*,1,1/3) + (\eta_1^0,\eta_1^-)(1,2,-1/2),
\end{eqnarray}
and singlets $N_{1,2,3}$ and $\chi$.  Let these be odd under Z$_2$ and the 
usual superfields be even.  Furthermore, let $N$ be odd under the usual 
matter parity and $\chi$ be even.  Then the usual $R$ parity of the Minimal 
Supersymmetric Standard Model (MSSM) is maintained together with the new 
Z$_2$.  Gauge-coupling unification of the MSSM is undisturbed if all the 
$\underline{5}$ and $\underline{5}^*$ particles are at the TeV scale. 
Proton decay is safe because $h$ and $h^c$ are odd under Z$_2$. The 
strong production of $h$ and $h^c$ at the LHC and their subsequent decays, 
such as $h \to d e^- \eta_2^+$ and $d e^+ \eta_2^-$, would yield 
same-sign dileptons plus quark jets plus missing energy, which is a 
possible unique signature of this model.

\subsection{Multipartite Dark Matter}
At least two out of the following three particles are dark-matter 
candidates: (1) the usual lightest neutralino of the MSSM with 
$(R,{\rm Z}_2) = (-,+)$, (2) the lightest exotic neutral particle with 
$(+,-)$, and (3) that with $(-,-)$.  The dark matter of the Universe may not 
be all the same, as most people have taken for granted!  For a general 
discussion, see Ref.~\cite{cmwy07}.

\subsection{Three-Loop Neutrino Mass}
Neutrino mass may also be obtained in three loops, with the addition of 
$N$ and a charged scalar $S^+_2$ which are odd under Z$_2$.  This 
model~\cite{knt03} also has a charged scalar $S_1^+$ which is even under 
Z$_2$.  It was the first proposal that $N$ could be dark matter.  
However, since $l N S_2^+$ is the only interaction involving $N$, it 
cannot be too weak to have the correct annihilation cross section for  
the observed dark-matter relic density.  This impliess generically large 
flavor-changing leptonic radiative decays, such as $\mu \to e \gamma$, 
and requires delicate fine tuning~\cite{kms06} to suppress.  In the 
one-loop scotogenic model~\cite{m06-2}, this constraint may be relaxed, 
because Re($\eta^0$) is available for dark matter.

Another three-loop model has been proposed~\cite{aks09} where the singlet 
$S_1^+$ is replaced by a a second Higgs doublet and a neutral singlet 
$\eta^0$ is added with odd Z$_2$.  In this case, $\eta^0$ (with a mass 
of 40 to 65 GeV) is a suitable dark-matter candidate.  Since its 
interactions with the other scalar particles are unconstrained, the 
desired relic abundance is easily obtained without requiring the 
$l N S^+$ coupling to be large, thus avoiding the problem of flavor 
violating leptonic interactions.  This model also allows for electroweak 
baryogenesis, coming from a first-order phase transition in the Higgs 
potential.  At the LHC, the $\eta^0$ of this model is hard to produce and 
detect because it is a singlet, whereas the Re$(\eta^0)$ and Im$(\eta^0)$ 
of the one-loop scotogenic model are produced by the $Z$ boson, with the 
subsequent decay Im$(\eta^0) \to$ Re$(\eta^0) l^+ l^-$ as a possible 
signature~\cite{cmr07}.

\section{Concluding Remarks}
\begin{itemize}
\item{With the application of the non-Abelian discrete symmetry A$_4$, a 
plausible theoretical understanding of the tribimaximal form of the neutrino 
mixing matrix has been achieved.}
\item{Seesaw variants at the TeV scale may allow this mechanism (in its 
inverse or linear manifestation) to be observable through the nonunitarity 
of the $3 \times 3$ neutrino mixing matrix, as well as flavor changing 
leptonic interactions.}
\item{Dark matter may be the origin of radiative neutrino mass.  This 
scotogenic mechanism may be implemented in a number of different models, 
and be observable also at the TeV scale.  A complete supersymmetric SU(5) 
version also exists with gauge-coupling unification.}
\item{Other recent topics in neutrino theory, such as Type III seesaw, using 
a Majorana fermion triplet $(\Sigma^+,\Sigma^0,\Sigma^-)$, and small Dirac 
and pseydo-Dirac neutrino masses, are not covered in this talk, but are being 
actively pursued.}
\item{Neutrino theory marches on, but what we really need are 
{\it corroborating data}!}
\end{itemize}

\begin{acknowledgments}
This work was supported in part by the U.~S.~Department of Energy under 
Grant No.~DE-FG03-94ER40837.
\end{acknowledgments}

\bigskip 

\begin{thebibliography}{99} 

\bibitem{c78} N. Cabibbo, Phys. Lett. {\bf B72}, 333 (1978).

\bibitem{w78} L. Wolfenstein, Phys. Rev. {\bf D18}, 958 (1978).

\bibitem{hps02} P. F. Harrison, D. H. Perkins, and W. G. Scott, Phys. Lett. 
{\bf B530}, 167 (2002).

\bibitem{m04} E. Ma, Phys. Rev. {\bf D70}, 031901 (2004).

\bibitem{m02} E. Ma, Mod. Phys. Lett. {\bf A17}, 2361 (2002).

\bibitem{es09} L. L. Everett and A. J. Stuart, arXiv:0812.1057 [hep-ph].

\bibitem{mr01} E. Ma and G. Rajasekaran, Phys. Rev. {\bf D64}, 113012 (2001).

\bibitem{m06-1} E. Ma, Mod. Phys. Lett. {\bf A21}, 2931 (2006).

\bibitem{af05} G. Altarelli and F. Feruglio, Nucl. Phys. {\bf B72}, 64 (2005).

\bibitem{m05} E. Ma, Phys. Rev. {\bf D72}, 037301 (2005).

\bibitem{l07} C. S. Lam, Phys. Lett. {\bf B656}, 193 (2007).

\bibitem{bhl08} A. Blum, C. Hagedorn, and M. Lindner, Phys. Rev. {\bf D77}, 
076004 (2008).

\bibitem{ahpz09} A. Atre, T. Han, S. Pascoli, and B. Zhang, JHEP {\bf 0905}, 
030 (2009).

\bibitem{ghj09} A. de Gouvea, W.-C. Huang, and J. Jenkins, arXiv:0906.1611 
[hep-ph].

\bibitem{ww83} D. Wyler and L. Wolfenstein, Nucl. Phys. {\bf B218}, 205 (1983).

\bibitem{mv86} R. N. Mohapatra and J. W. F. Valle, Phys. Rev. {\bf D34}, 
1642 (1986).

\bibitem{m87} E. Ma, Phys. Lett. {\bf B191}, 287 (1987).

\bibitem{m09-1} E. Ma, arXiv:0904.1580 [hep-ph].

\bibitem{bw90} W. Buchmuller and D. Wyler, Phys. Lett. {\bf B249}, 458 (1990).

\bibitem{p05} A. Pilaftsis, Phys. Rev. Lett. {\bf 95}, 081602 (2005).

\bibitem{ks07} J. Kersten and A. Yu. Smirnov, Phys. Rev. {\bf D76}, 073005 
(2007).

\bibitem{hm09} X.-G. He and E. Ma, arXiv:0907.2737 [hep-ph].

\bibitem{bm88} K. S. Babu and E. Ma, Phys. Rev. Lett. {\bf 61}, 674 (1988).

\bibitem{aa09} F. del Aguila and J. A. Aguilar-Saavedra, Nucl. Phys. 
{\bf B813}, 22 (2009).

\bibitem{k08} S. Khalil, J. Phys. {\bf G35}, 055001 (2008).

\bibitem{hkor08} K. Huitu, S. Khalil, H. Okada, and S. K. Rai, Phys. Rev. 
Lett. {\bf 101}, 181802 (2008).

\bibitem{ko09} S. Khalil and H. Okada, Phys. Rev. {\bf D79}, 083510 (2009).

\bibitem{adrs09} R. Allahverdi, B. Dutta, K. Richardson-McDaniel, and Y. 
Santoso, Phys. Rev. {\bf D79}, 075005 (2009).

\bibitem{kkr09} Y. Kajiyama, S. Khalil, and M. Raidal, Nucl. Phys. {\bf B820}, 
75 (2009).

\bibitem{bbms09} L. Basso, A. Belyaev, S. Moretti, and C. H. 
Shepard-Themistocleous, arXiv:0812.4313 [hep-ph].

\bibitem{bmt09} K. S. Babu, Y. Meng, and Z. Tavartkiladze, arXiv:0901.1044 
[hep-ph].

\bibitem{m09-2} E. Ma, Phys. Rev. {\bf D80}, 013013 (2009).

\bibitem{fhl09} P. Fileviez Perez, T. Han, and T. Li, arXiv:0907.4186 (2009).

\bibitem{m06-2} E. Ma, Phys. Rev. {\bf D73}, 077301 (2006).

\bibitem{dm78} N. G. Deshpande and E. Ma, Phys. Rev. {\bf D18}, 2574 (1978).

\bibitem{bhr06} R. Barbieri, L. J. Hall, and V. S. Rychkov, Phys. Rev. 
{\bf D74}, 015007 (2006).

\bibitem{m06-3} E. Ma, Ann. Fondation de Broglie {\bf 31}, 85 (2006).

\bibitem{di02} S. Davidson and A. Ibarra, Phys. Lett. {\bf B535}, 25 (2002).

\bibitem{m08} E. Ma, Phys. Lett. {\bf B659}, 885 (2008).

\bibitem{cmwy07} Q.-H. Cao, E. Ma, J. Wudka, and C.-P. Yuan, arXiv:0711.3881 
[hep-ph].

\bibitem{knt03} L. M. Krauss, S. Nasri, and M. Trodden, Phys. Rev. {\bf D67}, 
085002 (2003).

\bibitem{kms06} J. Kubo, E. Ma, and D. Suematsu, Phys. Lett. {\bf B642}, 18 
(2006).

\bibitem{aks09} M. Aoki, S. Kanemura, and O. Seto, Phys. Rev. Lett. {\bf 102}, 
051805 (2009).

\bibitem{cmr07} Q.-H. Cao, E. Ma, and G. Rajasekaran, Phys. Rev. {\bf D76}, 
095011 (2007).


\end{thebibliography}

\end{document}